\documentclass[aps,12pt]{revtex4}
\usepackage{epsfig}

\newcommand{\be}{\begin{equation}}
\newcommand{\ee}{\end{equation}}
\newcommand{\bea}{\begin{eqnarray}}
\newcommand{\eea}{\end{eqnarray}}
\newcommand{\bean}{\begin{eqnarray*}}
\newcommand{\eean}{\end{eqnarray*}}

\newcommand{\gapproxeq}{\lower
.7ex\hbox{$\;\stackrel{\textstyle >}{\sim}\;$}}
\newcommand{\lapproxeq}{\lower
.7ex\hbox{$\;\stackrel{\textstyle <}{\sim}\;$}}

\newcommand{\bc}{\begin{center}}
\newcommand{\ec}{\end{center}}
\newcommand{\btab}{\begin{tabular}}
\newcommand{\etab}{\end{tabular}}

\normalsize
\begin{document}

\title{\Large \bf On baryon-antibaryon coupling to two photons}

\author{Frank E. Close$^1$\footnote{\tt{e-mail: F.Close1@physics.ox.ac.uk}}
and Qiang Zhao$^2$\footnote{\tt{e-mail: Qiang.Zhao@surrey.ac.uk}}}

\affiliation{1) Department of Theoretical Physics, University of Oxford, 
Keble Rd., Oxford, OX1 3NP, United Kingdom}
\affiliation{2) Department of Physics, University of Surrey, 
Guildford, GU2 7XH, United Kingdom}


\begin{abstract}
We discuss recent claims that $p\bar{p} \to \gamma \gamma$
may be described by a generalized parton picture. 
We propose that quark-hadron duality 
provides a justification for the effective dominance of the ``handbag" diagram 
assumed in recent literature, and that handbag diagrams may dominate phenomena
in kinematic regions far more extensive than that might be expected from pQCD alone.

\end{abstract}

\maketitle

\section{Introduction}
Recently there has been the interesting 
proposal~\cite{freund,weiss,diehl-99,diehl,diehl-1}
 that exclusive proton-antiproton annihilation into two
photons can, under certain kinematic conditions, 
be described by a generalized partonic picture. As such,
this gives a new potential probe of 
non-forward parton distribution or
double distributions~\cite{ji,rad}.
It is argued~\cite{freund,weiss,diehl,diehl-1} 
that the two photons are emitted in the annihilation of
a single quark and antiquark (thus via the ``handbag" graph) 
whereby the process may be described by
a generalised parton picture analogous to the ``soft mechanism" 
in wide-angle real Compton
scattering. An essential feature of the arguments is 
that for on-shell photons at large $s$, the limited
space-like virtuality of the bound state wavefunctions 
constrains the active quark and antiquark to be ``fast",
i.e. have $x \to 1$. 

The purpose of this paper is to show that 
extension of recent work on quark-hadron duality~\cite{fcisgur,fcqz1,ijmv} may
help to justify some of the arguments that enable
phenomenology of the handbag diagram to be employed away from fully
asymptotic regime.

References~\cite{diehl-99,diehl} demonstrated that a ``soft-handbag" contribution (defined
by restricted transverse momenta and virtualities of quarks) factorises
into a soft hadronic matrix element and a partonic subprocess,
$\gamma \gamma \to q \bar{q}$ in the limit $x \to 1$ and large
$s,-t,-u$. However, there is no general proof that such soft-handbag topology
dominates for large (finite) $s$ though there are indirect hints of its importance
from phenomenological fits to data in e.g. $\gamma \gamma \to \pi\pi, K \bar{K}$
\cite{diehl}. The description for large intermediate $s$ implies that the spectators
possess $x \leq \Lambda_{QCD}^2/s$ and that the $x$ of the active quark is not literally
$\to 1$, but is in the region of $\sim 0.7$. It is plausible that soft wavefunctions can tolerate such values
of $x$. However, the factorization, which is reliable as $x \to 1$, has corrections
$\sim (1-x)^n$ which may become large as $x <1$.

In the region of $s$ that is in the hadronic continuum above the prominent resonances,
we argue that quark-hadron duality provides some justification for neglecting coherent
(``cats-ears" topology) that break factorization. It is also important to contrast the
soft handbag for finite $s$ and $x \neq 1$ from the ``hard" handbag for $s \to \infty$,
$x \to 1$ where quarks and gluons have large $p_T$ or virtuality. We make some brief comments
here in order to distinguish from the kinematics of duality and the ``soft" handbag, which
forms our main focus. Thus, in what
follows, we propose that 
duality should provide a justification 
for the handbag diagram dominance over a wide kinematic range. In particular, 
for $s$ above the resonance region, the presence of 
destructive interference among parity-even and parity-odd states~\cite{fcisgur}
in the coherent process for the $p\bar{p}$ annihilation indeed supports
the assumption of the handbag diagram dominance that underpins 
Refs.~\cite{freund,weiss,diehl-99,diehl,diehl-1} such that
 an effective incoherent parton interpretation
can be made, but integrated over $x$. There is no necessary dominance 
of the $x \to 1$
domain in this process.

 Thus for example, were it possible to compare the $p\bar{p}$ and
analogous $n\bar{n}$, the ratio of  
$\sum e_i^2(p)/\sum e_i^2(n)$ would be $3/2$.
This corresponds to a value of 1/2 
for the fragmentation parameter $\rho$~\cite{diehl-1}.

\section{Handbag dominance and $x \to 1$}

 Reference~\cite{freund,weiss} considers $p\bar{p} \to \gamma \gamma$ to be 
 Compton scattering in the 
crossed channel, where the exchanged system in the $t (u)$ 
channel consists of at least three quarks. At large momentum transfers 
such a
configuration should be strongly suppressed by the composite systems' 
wavefunctions. Hence
Refs.~\cite{freund,weiss} argue that the most efficient way of accommodating 
a large momentum
transfer is via the handbag diagram (where the $p\bar{p}$ system makes
a transition to a $q\bar{q}$ pair by exchanging a virtual ``diquark" system
whose spacelike virtuality is limited by the bound-state wavefunction). 
The $q\bar{q}$ then annihilate into two photons by exchanging a highly virtual
quark/antiquark.

\begin{figure}
\begin{center}
\epsfig{file=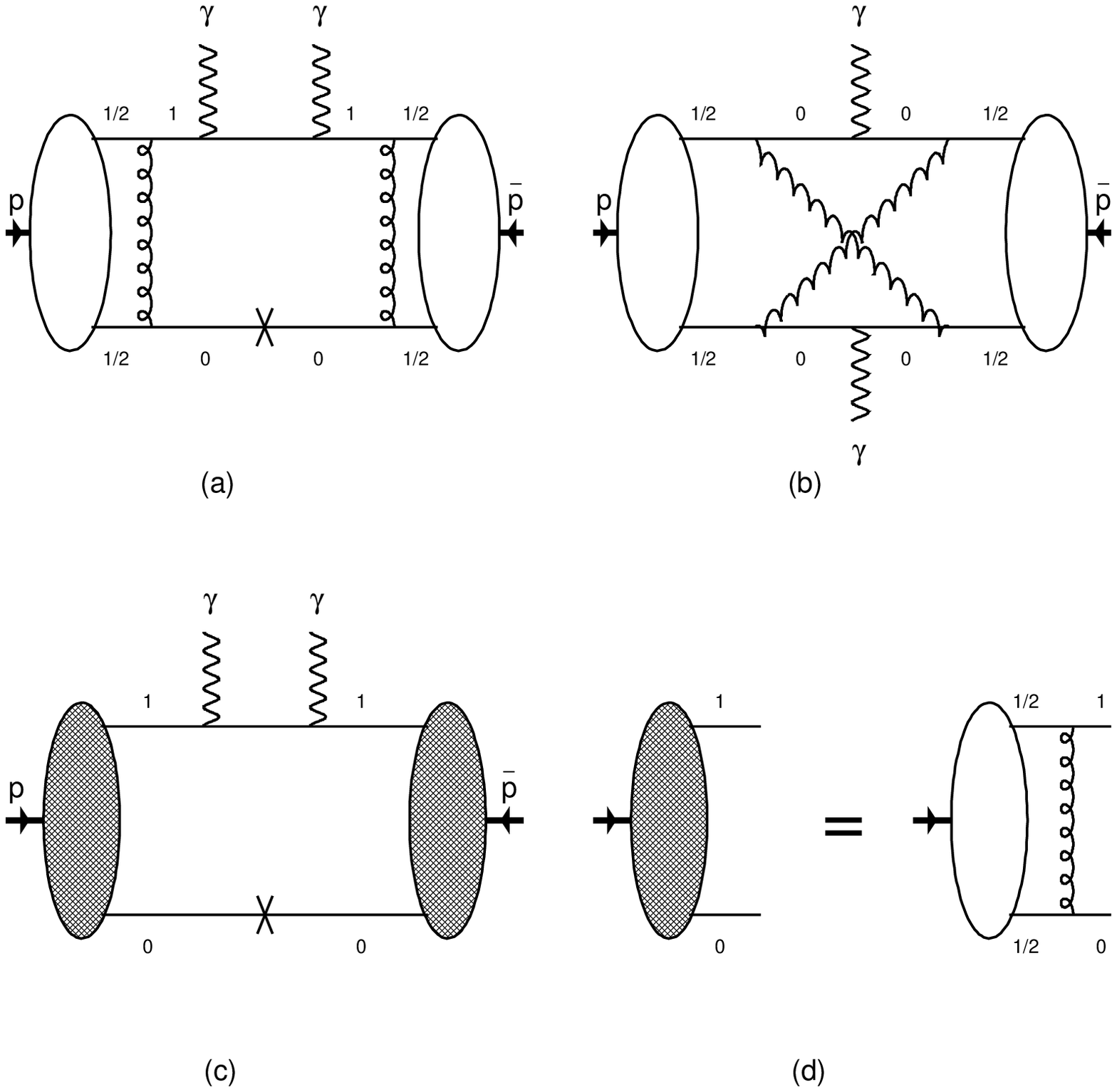,width=12cm,height=12cm}
\caption{Schematic diagrams for (a) incoherent photon emission and
(b) coherent photon emission; Fig. (c) is the ``handbag" which is the 
same order as (a); Fig. (d) illustrates the kinematical 
access of the soft ``handbag" via gluon exchange.
 }
\protect\label{fig-1}
\end{center}
\end{figure}

There is an essential difference between the annihilation process and the
forward Compton scattering that underpins deep inelastic structure functions.
In the latter, the spectator (``diquark", etc) system is effectively passive,
the active quark probability being factorized out and probed by the highly
virtual photon(s). Contrast this with 
the exclusive annihilation process, $p\bar{p} \to \gamma \gamma$, where
one may annihilate a $q\bar{q}$ by $q\bar{q} \to \gamma \gamma$, but must also 
annihilate the spectators without the emission of further radiation 
(be it photons, gluons or hadrons). It is this constraint in the strict limit
$s \to \infty$ that effectively 
requires the spectators to have null
energy-momentum four-vectors, and hence the active $q\bar{q}$ to carry the full
momentum of the beams (what Refs.~\cite{freund,weiss} refer to as ``fast" quarks
and which Ref.~\cite{diehl} use as their driving assumption in developing
phenomenology). However, there is no immediate argument to support
 the handbag diagram {\it dominance}. For $s \to \infty$, where pQCD applies, 
one can see below that the (neglected) coherent or higher-twist
contribution of Fig.~\ref{fig-1}b, is the
same order of magnitude as the ``handbag" diagram (Fig.~\ref{fig-1}a). We briefly review
this to distinguish it from the large intermediate $s$, where the ``soft-handbag"
dominance may apply, and turn to this in section III.

The nucleon wavefunction restricts the spacelike virtuality
of partons to be small~\cite{freund}.
Production of a single real photon 
at large transverse momentum in parton-antiparton annihilation,
then requires a rather singular kinematics, with 
$p_i = (|p_T|, p_T, {\cal O}(\Lambda_{QCD}/\sqrt{s}))$ 
for each parton (hence each $x \to 0$), where $\Lambda_{QCD}$ is the 
typical QCD energy scale.
However, such a singular kinematics in principle can occur.  
A coherent amplitude, as in Fig.~\ref{fig-1}b, requires transfer of momentum
so that at large $s$
the annihilating constituents all have $x \to 0$. This can
be achieved by means of gluon exchange but at the expense of suppression
due to the large momentum flows and powers of $\alpha_s$. 
(We illustrate this for a two-body system 
but the argument generalizes). The superscripts in Fig.~\ref{fig-1} denote 
the longitudinal momentum fractions;
the most favoured configuration at the hadron vertex is for the 
constituents to share their momenta (see e.g. Refs.~\cite{fcqz1,ijmv}).

The restricted kinematics ($x\to 1$) is unable to guarantee the dominance
of the handbag diagram illustrated by Fig.~\ref{fig-1}c. 
The configuration in Fig.~\ref{fig-1}c, 
is also kinematically singular, in that the parton
probabilities vanish in the strict $x \to 1$ limit, and the
spectators must be null in order for their annihilation to contribute nothing.
 One thus needs to consider how this extreme
configuration arose if, as seems natural, the preferred wavefunction
(denoted by the non-shaded ovals in Figs. 1a,b,d)
has the partons with symmetric configuration 
$x \sim 1/2$ (for this pedagogic example 
of a two body system ~\cite{fcqz1}).
If one allows gluon exchanges as in Fig.~\ref{fig-1}a to achieve this singular
kinematics, it would 
lead to a configuration that is generally not suppressed 
relative to the
one in Fig.~\ref{fig-1}b.
Figure~\ref{fig-1}d then shows 
how the intrinsic symmetric (in $x$) wavefunction 
becomes highly asymmetric.
This causes Fig.~\ref{fig-1}c to be the same order 
in $\alpha_s$ and momentum flow as Fig.~\ref{fig-1}b.

Interestingly, 
for $q^2 > 0$ timelike processes e.g. $\gamma \gamma \to \pi \pi$, 
coherent diagrams such as those in Fig.~\ref{fig-1}b
are anticipated in pQCD~\cite{brodsky}. The relative strengths 
of various meson-pair
production processes at finite $s$, such
as $\sigma(\gamma\gamma \to\pi^0 \pi^0)/\sigma(\gamma\gamma\to \pi^+ \pi^-)$ 
do not fit well with these predictions in detail. Indeed,
phenomenology seems compatible with the dominance of handbag diagrams at
intermediate $s$. This is the subject of soft-handbag
dominance~\cite{freund,diehl-99,diehl,diehl-1,weiss} to which we now turn.

\section{Soft handbag and duality at intermediate energies}

We shall now propose that quark-hadron duality
may provide an explanation of the phenomenological success
of data analyses based on the assumed dominance of the incoherent, 
or ``handbag" diagrams
at intermediate $s$, and help
underpin recent developments based thereon 
~\cite{weiss,diehl,diehl-1,freund}.

In quark-hadron duality the 
incoherent property of the handbag diagram survives in practice
even though the kinematic arguments of pQCD 
do not necessarily apply~\cite{fcgilman,fcisgur}.
The underlying dynamics lead to the effective
dominance of the incoherent diagrams when a suitable 
averaging has taken place~\cite{fcisgur}. 
We first illustrate some empirical examples 
and then apply the
arguments to the $p\bar{p} \to \gamma \gamma$ 
process~\cite{weiss,diehl,diehl-1,freund}.

In the deep inelastic structure functions at large $q^2$ one has
$F_2^n(x)/F_2^p(x) = \sum e_i^2(n)/\sum e_i^2(p)$ 
in kinematic circumstances where pQCD
supports the incoherent dominance, 
and where any $\sum_{i\ne j} e_i e_j$ contributions are higher twist
and thereby suppressed. However,
incoherent contributions proportional to $\sum_i e_i^2$ appear to control
the cross-section $ratios$ of Compton scattering even at low $q^2$ 
where the kinematic conditions are such that $\sum_{i\ne j} e_i e_j$ 
contributions 
would be anticipated. As an extreme example, consider real photons, 
where the non-diffractive contribution to 
$\sigma_{tot}(\gamma n)/\sigma_{tot}(\gamma p)$ is empirically 
$\sim 2/3 \equiv (2e^2_d + e^2_u)/(2e^2_u + e^2_d)$,
even though there is no pQCD support for such a relation when $q^2= 0$.
The suppression of the $\sum_{i\ne j} e_i e_j$ contributions 
in this case is a result of duality in 
$\mbox{Im}(\gamma N \to \gamma N (t=0))$
and the constraint that there are no exotic exchanges 
in the crossed $(t)$ channel~\cite{fcgilman}. In a pedagogic model
Reference~\cite{fcisgur} showed how this can arise. Furthermore,
this model has been shown to realise scaling in the structure
functions (``Bloom-Gilman duality")~\cite{ijmv} and to lead to 
a factorization of the non-forward Compton
scattering~\cite{fcqz1}. 
The duality holds for real photons as $t\to 0$, and the factorisation
at least for the non-forward Compton scattering of real photons.
It is the latter that crosses to $p\bar{p} \to \gamma\gamma$, 
whereby we propose that this may also justify 
the dominance of the incoherent process ($\sim\sum_i e_i^2$) 
in $p\bar{p} \to \gamma \gamma$, but without 
any restriction to $x \to 1$.

In $\mbox{Im}(\gamma N \to \gamma N (t=0))$ the excitation of coherent 
intermediate resonances includes
states of positive and of negative parity. 
These add constructively in the $e_i^2$ contributions
but are destructive in the $e_ie_j$ terms. 
This causes a duality between the averaged resonance
excitation on the one hand, 
and the smooth high-energy behaviour on the other hand~\cite{fcisgur}. 
As a consequence one obtains
the empirical result for the ratio of the non-diffractive pieces of
$\sigma_{tot}(\gamma n)/\sigma_{tot}(\gamma p)$ even though 
there is no pQCD reason for the dominance
of the incoherent terms.

This picture has been extended to the spin-averaged non-forward Compton 
amplitude~\cite{fcqz1}, where scaling and factorization properties arise.
In particular, the same effective
dominance of the $\sum_i e_i^2$ terms arose. 
It implies for $\gamma(q) A \to
\gamma(k) A$ with $t \equiv (k-q)^2$,
the generalized factorization for the non-forward proton structure function
or double distribution function~\cite{fcqz1}
is
\be
\label{factor}
F_2(x,\xi,t)=\sum_i e_i^2 \frac{(x-\xi)(x+\xi)}{x^2} F_2(x) F_{el}(t) \ ,
\ee
where $F_{el}(t)$ is the elastic form factor satisfying $F_{el}(t=0)=1$.
In the particular limit
of $\xi\to 0$ and $-t/Q^2<< 1$, this factorization also satisfies 
Ji and Radyushkin's sum rule~\cite{ji,rad}.

The analogous analysis can be applied to
the crossed channel, $\gamma \gamma \to M \bar{M}$, 
where $M$ refers to a two-body ``meson". The
essential results 
can then be generalised to $\gamma \gamma \to p\bar{p}$ 
of Ref.~\cite{weiss,diehl,diehl-1,freund}. The most general consequence
is again that terms proportional to $\sum_{i\ne j} e_ie_j$ 
are suppressed by destructive
interference (in the $t$-channel). The corresponding effect is 
that the dominant process in the
$s$-channel is $\gamma \gamma \to M(q\bar{q}) \to p\bar{p}$, 
where the intermediate meson states $M$
have been summed over, and all ``exotic" $qq\bar{q}\bar{q}$ 
intermediate states been suppressed. This underpins the
phenomenology of Ref.~\cite{diehl}. In particular, such arguments also immediately imply
$\sigma(\gamma\gamma\to\pi^+\pi^-)\simeq\sigma(\gamma\gamma\to\pi^0\pi^0)$
in the hadronic continuum.

We can compare our distribution function with 
that used in Ref.~\cite{freund}. There, 
a factorization ansatz was made for the double distribution function:
$F_{\alpha}(x,\alpha;s) = f_{\alpha}(x) h_{\alpha}(x,\alpha) 
S_{\alpha}(x,\alpha,s)$,
separating the soft and hard contributions with $S_{\alpha}(x,\alpha,s=0) =1$ 
and $\int h(x,\alpha) d\alpha = 1$. Furthermore, it was assumed that
$h_{\alpha}(x,\alpha) \equiv \delta(\alpha)$.

If we now impose an analogous set of ansatz on our form [Eq.~(\ref{factor})]:
$F(x) h(x,\xi) S(x,\xi,t)$ with $S(x,\xi,t=0) =1$, 
 $\int h(x,\xi) d\xi = 1$ and
$h(x,\xi) \equiv \delta(\xi)$, and then cross {\it t} to {\it s}
channel, we obtain analogues of their Eqs. (6) and (7)~\cite{freund}. 
Namely, the   
factorization satisfies 
\be
F_2(x,\xi\to 0,s) = \sum_i e_i^2 F_2(x) F_{el}(s) 
= \sum_i e_i^2 x q(x)F_{el}(s) \ ,
\ee
and 
\be
\label{dvcs-fact}
\frac 12 \int_{\xi=-1+|x|}^{1-|x|} 
 d\xi \frac{F_2(x,\xi,s=0)}{x} = 
 \sum_i e_i^2 [\theta(x) q(x) -\theta(-x) \overline{q}(-x)] \ ,
\ee
where $\xi\to 0$ forces us throw away the term proportional to $\xi^2$;
$q(x)$ is 
the unpolarized quark (antiquark) distribution function,
and $\theta(x)$ is the step function. 
The factor 1/2 comes from 
the re-definition of $\xi\equiv x_{bj}^{in}-x_{bj}^{fin}$ in contrast 
with $\xi\equiv (x_{bj}^{in}-x_{bj}^{fin})/2$ in Ref.~\cite{fcqz1},
and the relation
$-\theta(x)F_2(x)+\theta(-x)F_2(-x)=0$ according to the momentum sum rule
has been used.
Taking this factorization scheme, the integration ranges for $x$ and 
$\xi$ become $-1<x<1$ and $-1+|x|<\xi < 1-|x|$, which are consistent
with the convention used in the literature~\cite{weiss,freund}.

\begin{figure}
\begin{center}
\epsfig{file=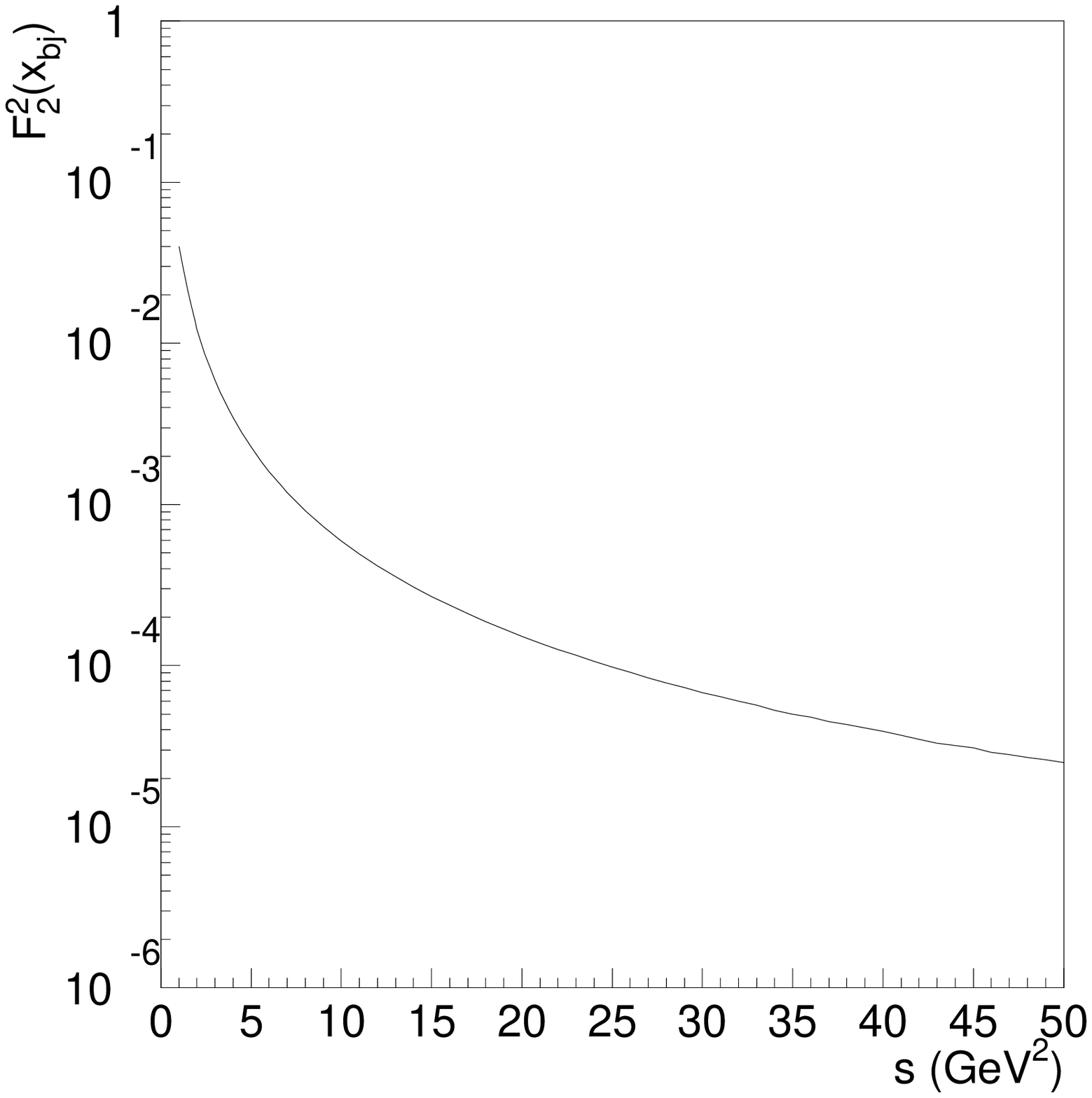,width=12cm,height=12cm} 
\caption{Squared form factor for $p\bar{p}\to \gamma\gamma$ 
predicted based on duality.}
\protect\label{fig-2}
\end{center}
\end{figure}

Our expression, which was generalized to the Compton scattering on the proton, 
then leads directly to 
the vector form factor for the proton defined in Ref.~\cite{freund}
for the $p\bar{p}\to \gamma\gamma$:
$R_V(s)\sim \sum_i e_i^2 F_{el}(s)$.
Although the generalization of the factorization to the physical nucleon
scattering has not been justified, we find that the above 
analytical expression from the factorization is rather interesting
in light of the numerical results of Ref.~\cite{freund,weiss}.
Given a dipole feature to the elastic form factor 
$F_{el}(s)=1/(1+s/\lambda^2)^2$, we plot the $R_V(s)$ in Fig.~\ref{fig-2}, 
which exhibits a similar feature as found in Ref.~\cite{weiss,freund}.
Meanwhile, one consequence is that the cross section ratios 
between the $p\bar{p}$ and 
$n\bar{n}$ annihilations are governed by the constituent charges, 
namely, 
$\sigma(p\bar{p})/\sigma(n\bar{n}) = (2e_u^2 + e_d^2)/(2e_d^2 + e_u^2)$. 
There is no restriction to specific extremes of $x$, nor is there a
freedom of fragmentation parameter $\rho$.

\section{Summary}

In the spirit of Refs.~\cite{fcgilman,fcisgur} 
as illustrated for DVCS and wide angle Compton scattering (WACS)
in Ref.~\cite{fcqz1},
one can argue for the effective dominance of the incoherent 
 diagrams, and the
ability to interpret wide-angle $p\bar{p} \to \gamma \gamma$ 
by non-forward distributions, as in Refs.~\cite{weiss,diehl,diehl-1,freund}.
There is,
however, no dynamical reason to require the dominance of $x \to 1$ 
configurations.

Meanwhile, note the link between the effective destruction 
of $\sum_{i\ne j} e_i e_j$ terms and
the absence of exotic exchanges in the crossed 
channel~\cite{fcgilman,fcisgur}. This
necessarily imposes constraints on the relative magnitudes 
of $\gamma \gamma \to A\bar{A}$.
In particular, under the kinematic circumstances 
of Refs.~\cite{diehl,brodsky}
$\sigma(\gamma \gamma \to \pi^+\pi^-)/\sigma(\gamma\gamma\to \pi^0\pi^0 ) = 1$. These results can be generalized to 
physical baryons and give the relations for listed 
in Ref.~\cite{diehl}, but where 
the above arguments would imply that $\rho$ 
in Eq. (43) of ~\cite{diehl} simply counts the relative number 
of $d$ and $u$ valence quarks in a proton. Assuming 
SU(3) flavour symmetry, one thus would
have $\rho =1/2$ in Eq. (43) of Ref.~\cite{diehl-1}.

In summary, duality enables us to generalise the results of 
Refs.~\cite{weiss,diehl,diehl-1,freund}.
In regions where pQCD is able to support ``local" dominance
of leading twist ``handbag" diagrams, quark-hadron duality shows that
such diagrams dominate certain ratios of cross sections 
if averaged over $s$ such that
direct channel resonances of opposite symmetry types destructively cancel. 
As $s \to \infty$
where the density of states is high, this is expected to happen locally, 
and the results of pQCD arise. 
At the other extreme, namely the region where individual
resonances dominate locally, 
cancelling only when the resonance region 
is suitably averaged~\cite{fcisgur,ijmv},  
such cancellations will not arise and the coherent effects dominate.
At intermediate $s$, as here, where the density of resonances rapidly
saturates, the cancellations are rather local and the handbag dominance
becomes effective. Thus in summary,
we support the hypothesis  
that the incoherent probability description
is valid. In particular, there is no necessary restriction 
to the $x \to 1$ limit. 

It is also worth noting that 
our model is manifestly symmetric by construction and
implicitly has $\rho =1/2$~\cite{diehl-1} and hence 
the ratio of $p\bar{p}$ to $n\bar{n}$ is 3/2. 
A realistic picture would
require extension to include spin and single gluon exchange, or symmetry
breaking effects. However, this goes beyond our immediate aims which were to
give some underpinning of the arguments for application of handbag
phenomenology to $p\bar{p} \to \gamma\gamma$.

\section*{Acknowledgements}

We thank P. Kroll, C. Weiss and A.V. Radyushkin for helpful discussions 
about their work. 
This work is supported, in part, by grants from the Particle Physics and
Astronomy Research Council,  and the
EU-TMR program ``Eurodafne'', NCT98-0169,
and the U.K. Engineering and Physical 
Sciences Research Council (Grant No. GR/M82141).


\begin{thebibliography}{99}
%
\bibitem{freund} A. Freund, A.V. Radyushkin, A. Sch\"afer, and C. Weiss, 
	hep-ph/0208061.
%
\bibitem{weiss} C. Weiss, hep-ph/0206295.
%
\bibitem{diehl-99} M. Diehl, T. Feldmann, R. Jakob, and P. Kroll,
	Eur. Phys. J. C {\bf 8}, 409 (1999).
%
\bibitem{diehl} M. Diehl, P. Kroll, and C. Vogt, 
	Phys. Lett. {\bf B 532}, 99 (2002).
%
\bibitem{diehl-1} M. Diehl, P. Kroll, and C. Vogt, 
	hep-ph/0206288.
%
%
\bibitem{ji} X. Ji, 
	Phys. Rev. Lett. {\bf 78}, 610 (1997);
	Phys. Rev. D {\bf 55}, 7114 (1997);
	J. Phys. G. {\bf 24}, 1181 (1998).
%
\bibitem{rad} A. Radyushkin, 
	Phys. Rev. D {\bf 59}, 014030 (1999); 
	Phys. Lett. {\bf B449}, 81 (1999).
%
\bibitem{fcisgur} F.E. Close and N. Isgur, 
	Phys. Lett. {\bf B 509}, 81 (2001).
%
\bibitem{fcqz1} F.E. Close and Q. Zhao, 
	Phys. Rev. D {\bf 66}, 054001 (2002), hep-ph/0202181.
%
\bibitem{ijmv} N. Isgur, S. Jeschonnek, W. Melnitchouk, and J.W. Van Orden,
	Phys. Rev. D {\bf 64}, 054005 (2001).
%
\bibitem{brodsky} S.J. Brodsky and G.P. Lepage, 
	Phys. Rev. D {\bf 22}, 2157 (1980).
%
\bibitem{fcgilman} F.E. Close and F.J. Gilman, 
	Phys. Lett. {\bf B38}, 541 (1972); 
	Phys. Rev. D {\bf 7}, 2258 (1973).
%
\end{thebibliography}
\end{document}